\documentclass[preprint]{aastex}

\newcommand{\hi}{{H~{$\scriptstyle {\rm I}$}}}

\begin{document}

\title{High Velocity Cloud Complex H: A Satellite of the Milky Way in a 
Retrograde Orbit? }

%% revised version 15 May 2003 -- revised Fig 1 and captions to 1 & 3.
 
\author{Felix J. Lockman}
\affil{National Radio Astronomy Observatory
\footnote{The National Radio Astronomy Observatory is a facility of the
National Science Foundation operated under cooperative agreement with
Associated Universities, Inc.}, P.O. Box 2, Green Bank, WV, 24944; 
 jlockman@nrao.edu}
\authoraddr{P.O. Box 2, Green Bank, WV, 24944}

\begin{abstract} 

Observations with the Green Bank Telescope 
of 21cm \hi\ emission from the high-velocity cloud 
Complex H suggest that it is interacting with the Milky Way.  
A model in which the cloud is a satellite of the Galaxy in an 
inclined, retrograde circular orbit 
reproduces both the cloud's average velocity and its 
velocity gradient with 
latitude.  The  model places Complex H at $R = 33 \pm 9$ kpc 
from the Galactic Center  on a retrograde
  orbit inclined $\approx 45\arcdeg$ to the 
Galactic plane.    At this location it has a mass in \hi\  of
 $> 6 \times 10^6 \ M_{\Sun}$ and dimensions of 
at least $10 \times 5$ kpc. 
Some of the diffuse \hi\ associated with the cloud has apparently been 
decelerated by interaction with Galactic gas.   Complex H has 
similarities to the dwarf irregular galaxy Leo A and to 
 some compact high-velocity clouds, and has an internal structure 
nearly identical to  parts of the Magellanic Stream, with a pressure 
$P/k \approx 100$ cm$^{-3}$ K. 
\end{abstract}
 
\keywords{Galaxy:halo --- Galaxy: structure --- ISM:clouds -- Radio Lines:ISM}

\section{Introduction}
The current model for galaxy formation is one of 
hierarchical aggregation \citep{freeman02}, and in this 
context the problem  of the nature of high-velocity \hi\ clouds 
has attracted renewed attention \citep{blitz}.  
The high-velocity \hi\ cloud ``Complex H'', first discovered 
by \citet{dieter} and \citet{huls71} and shown in Figure 1, 
 covers more than 100 square degrees  at very low 
 Galactic latitude.  Its brighter components have a velocity with
respect to the Local Standard of Rest  of 
$V_{LSR} \approx -200$ km s$^{-1}$, which, in this direction, 
is at least 35 km s$^{-1}$ more negative than is 
allowed for an object in circular Galactic rotation  at any distance, 
and  about 100 km s$^{-1}$ more 
negative than the observed minimum velocity from \hi\  in 
the Galactic disk.  It lies in the Galactic plane, so the anomalous 
motion cannot be attributed to infall.

\citet{wakker98} show that  Complex H 
must be  more than 5 kpc from the Sun, 
which eliminates a possible association with the Local Bubble, 
and they dismiss several other ``local'' models on the
grounds that the cloud does not seem to show strong interaction with 
Galactic gas.   \citet{morras}, however, 
note that Complex H overlaps in position with a large \hi\ shell at 
permitted velocities, and propose 
 that it created the shell as it  penetrated the Galactic disk 
about 22 kpc from the Galactic Center.  
This interpretation has been challenged by  \citet{blitz},  
 who cite a lack of ancillary evidence for 
 so energetic an encounter and, instead, hold Complex H to be an 
important link in a chain of evidence which 
leads them to conclude that most high-velocity clouds 
are at intergalactic rather than interstellar distances.

In fact, the basic conclusion that the velocity of 
Complex H cannot arise from circular rotation about the Galaxy is  correct 
only for prograde orbits.  An LSR velocity of $-200$ km s$^{-1}$ at 
$\ell = 130\arcdeg$ can be achieved by an object 
with $V_{\theta}/R \approx -5$ km s$^{-1}$ kpc$^{-1}$, 
where R is the distance from 
the Galactic center,  $V_{\theta}$ is the azimuthal velocity, 
and the negative sign indicates that the 
motion is against the sense of normal Galactic rotation.  
This Letter presents 
 21cm \hi\ observations of Complex H made with the Green Bank 
Telescope which show that there is  \hi\ at velocities 
between the cloud and the disk, confirming previous suspicions
 that it is interacting with the Milky Way.  
The new data are consistent with a  model 
in which this object is a satellite of the Milky Way 
in an inclined, retrograde orbit, whose outermost
layers are currently being stripped away in its encounter with the Galaxy.

\section{Complex H and the Galactic Disk}

Figure 2 shows a  cut through Complex H at  $\ell=130\fdg5$ 
 made from the Leiden-Dwingeloo (LD) 21cm \hi\ survey data 
 \citep{hartmann}.  The cloud 
divides into two distinct sections: a bright ``core''
 which lies at $b \lesssim 5\arcdeg$, 
and a fainter, more diffuse ``tail'' at $b > 5\arcdeg$.  
Both the core and the tail move to more positive 
velocities away from the cloud apex near $b \approx 4\arcdeg$, 
 the tail reaching velocities allowed by Galactic rotation 
at $b \approx 10\arcdeg$.   This same general structure 
is seen  at many longitudes 
and was apparent in the earliest observations \citep{dieter,huls75}.
There is also a hint of  faint \hi\ 
emission at velocities between that of the cloud core and 
the Galactic disk. 

Part of Complex H at $130\fdg5 \leq \ell \leq 132\fdg5$ and 
$-2\arcdeg < b < +8\arcdeg$ was observed in the 21cm line with the 
Robert C. Byrd Green Bank Telescope  (GBT) in August 2002 
 at $9\arcmin$ angular resolution with $3\arcmin$ spacing 
between spectra.  Each spectrum covers 
 $  -300 \leq V_{LSR} \leq +200$  km s$^{-1}$ at 
 1.25 km s$^{-1}$  velocity resolution. 
The $1\sigma$ noise in a  channel is 0.16 K.  The equipment and 
 observing methods are identical to those described 
elsewhere  \citep{fjl2002}.

Figure 3 shows  the GBT data at $\ell = 130\fdg6$.  
About $40\%$ of the spectra through the core of the Complex 
have two line components, one broad and one narrow, 
with a median $\Delta v $ (FWHM) of 8 and 28 km s$^{-1}$.  
In contrast, lines from the tail are almost uniformly broad 
with a median $\Delta v = 34$ km s$^{-1}$.  The median width of 
all lines from the core is 22 km s$^{-1}$, while almost every 
spectrum through the tail has $\Delta v > 22$ km s$^{-1}$.  
  This suggests that
the division between core and tail is not arbitrary, but is related
to the nature of  the Complex. 
Fig.~3 also shows diffuse \hi\ emission filling much of the 
 gap in velocity between the Complex 
and the Galactic disk, strong evidence for an 
interaction between the Complex and the Milky Way.  
The Complex is not nearly as  isolated from the Galaxy 
as it seemed from earlier observations.  

\section{A Model for Complex H}

 In the model proposed here  
Complex H has a core  which  moves like a solid body  
 on a circular orbit, and a tail of diffuse, decelerated gas resulting from 
its interaction with the Milky Way.  
The velocity of an object  projected to the LSR can be written 

\begin{equation}
 V_{LSR} =  \left[ R_0\  \sin(\ell)\ \lbrace  {V_{\theta}\over R}
 - {V_0\over R_0 } \rbrace \ -V_R\ \cos(\ell+\theta) \right]\cos(b)\
 + V_z \sin(b), 
\end{equation}
in  an $R,\theta,z$  polar coordinate system 
centered on the Galactic center, 
where $\theta$ runs in the direction of Galactic rotation 
from the Sun-center line, and $z$ is toward the North Galactic Pole.  
The standard Galactic constants, $R_0 = 8.5$ kpc, $V_0 = 220$ 
km s$^{-1}$ are used throughout.
If Complex H moves in a circular orbit ($V_R = 0$) about the Galactic center 
 with a  velocity of magnitude $V_c$, 
at $b = 0\arcdeg$ it will have  components 
$V_{\theta} = V_c\ \cos(i)$ and $ V_z = V_c\  \sin(i)$, where
$i$ is the inclination of the orbit to the Galactic plane: 
$i = 0\arcdeg$ is prograde in the Galactic plane, and 
$i = 90\arcdeg$ is toward the North Galactic Pole. 

The data constrain $V_z$ and $V_\theta$.  At low 
latitudes $\cos(b) \approx 1$ and $\sin(b) \approx b$, 
so the gradient $dV_{LSR}/db \approx V_z$.  
The observed slope in $V_{LSR}$ of the core is 
 about --3 km s$^{-1}$ per degree of latitude 
giving   $V_z = V_c \sin(i) = -170$ km s$^{-1}$. 
Constraints on $V_\theta/R$ come at  $b=0\arcdeg$, 
where the second term of eq.~1 is zero.  At 
$\ell,b = 130\fdg5+0\arcdeg$ the core has   
$V_{LSR} = -195 $ km s$^{-1}$ implying that 
 $ V_{\theta} /R = -4.3$ 
 km s$^{-1}$ kpc$^{-1}$, or $R = - 0.23 \ V_c\ \cos(i)$ kpc. 
As the Galactic rotation curve is likely to be nearly flat 
 to several hundred kpc from the Galactic center 
\citep{zaritsky}, setting $V_c = V_0 = 220$ km s$^{-1}$ 
 yields the  orbital inclination $i = 230\arcdeg$ and the 
distance from the Galactic center  $R = 33 $ kpc.
 Complex H thus appears to be moving to greater 
longitude and to negative latitude at an angle $\approx 45\arcdeg$ 
with respect to the Galactic plane. The direction of motion is
drawn in Fig.~1.

These considerations are used to produce the 
 simulation displayed in Figure 4.  Here it is assumed that the 
Galaxy has a flat rotation curve with $V_c = 220$ km s$^{-1}$ and 
Complex H has two components, both with $\Delta v = 22$  
km s$^{-1}$.  The core follows the retrograde, inclined, 
orbit $V_c = 220$ km s$^{-1}$, 
$i  = 230\arcdeg$, at a distance $R=33 $ kpc from the Galactic 
center.  The diffuse gas is 
slightly more than twice the size of the core and extends 
to higher latitude;    its velocity 
varies linearly between the velocity of the core where they meet, and  
normal Galactic disk rotation,
 $V_\theta=V_0$ and $V_z=0$,  at its outer edge.  
This simulation is not intended to fit the structure of Complex  H in detail, 
 but simply  to check the parameters  of 
  the proposed model. On this criterion the model does well. 
 The simulated core has the correct velocity and gradient, and the 
lagging diffuse gas  produces a credible tail.

Reasonable fits to the GBT data can be found for models 
in the range $V_z = -170 \pm 30 $ km s$^{-1}$, 
implying an approximate range of parameters 
$i = 230 \pm 10\arcdeg$ and  $R = 33 \pm 9$ kpc.  
The orbit crosses the Galactic plane at $\theta = 38\pm 5\arcdeg$ 
from the Sun-center line in the direction of Galactic rotation 
at a distance from the Sun $r = 27 \pm 9 $ kpc. 
If the Galactic circular velocity declines beyond the Sun as is 
suggested by some evidence \citep{zaritsky}, a somewhat larger 
inclination would be derived, along with a  smaller value of R.  

\section{Implied Properties of Complex H}

At a distance from the Sun $r=27$ kpc  the total \hi\ mass  
over $V_{LSR} \leq -175$ km s$^{-1}$  (the shaded area of Fig.~1) 
 is  $ 6.4 \times 10^6 \ M_{\sun}$.  About half of this 
resides in the 
bright core at $\ell \geq 127\arcdeg, b\leq +5\arcdeg$ while about 
$10\%$ is in the diffuse tail at $b > 5\arcdeg$. 
There is certainly \hi\ associated with the Complex 
at more positive velocities (e.g., that shown in the Fig. 1 contours), 
but the  restrictions adopted here avoid 
possible confusion with unrelated material and provide 
 a firm  lower limit to the total mass.  
The size of the core in the $\ell$ and $b$ directions is $\approx 2 \times 3$ 
kpc, while the total extent of the cloud is 
at least $10 \times 5$ kpc.    
The dynamical mass needed to bind a spherical cloud of radius $r_c$ 
and line width $\Delta v$ is $M_{dyn} \approx r_c \Delta v^2 /G$, 
where $G$ is the gravitational constant.  The core of Complex H, 
with $r_c \sim 1$ kpc and median $\Delta v = 22 $ km s$^{-1}$,  
has $M_{dyn} \approx 10^8 M_{\Sun}$, while the core \hi\ 
mass is only $3 \times 10^6 M_{\Sun}$.  Complex H is unlikely to  be 
gravitationally bound by its \hi\ alone.

The two-phase structure of the \hi\ lines from the core, 
 also seen in other high-velocity clouds, suggests that it is 
in pressure equilibrium with a surrounding, low-density, 
  medium \citep{cram,ferrara94}.  
There is a group of  \hi\ features in the Complex at $130\arcdeg+1\arcdeg$ 
 that have a characteristic angular size of $4\arcmin$ \citep{WS91}. 
At $r=27$ kpc they would have a diameter of 30 pc and a typical 
$ \langle n \rangle \approx 2$ cm$^{-3}$. 
Their temperature is likely 50 K \citep{WakkVS} implying a 
 pressure  $P/k \equiv nT \approx 100$ cm$^{-3}$ K.  
An identical pressure is derived from the GBT data, using the 
median core $\Delta v$ of 22 km s$^{-1}$ to limit the temperature, 
and the median core $N_{HI}$ of  $   6 \times 10^{19}$ cm$^{-2}$ 
over a size of 2 kpc for an average volume density. 

 Toward Complex H  there is likely 
$> 10^{22}$ cm$^{-2}$ of foreground \hi.  
Studies of the region using the 2MASS survey 
have been complicated  by strong differential reddening  
$(0.6 \lesssim $ E(B-V) $\lesssim 3)$, 
but no obvious stellar features attributable to Complex H are found 
in differenced, dereddened (J - K$_s$, K$_s$)$_o$ 
color-magnitude diagrams compared with nearby control fields 
(S. Majewski \& M. Skrutskie, private communication).  
There is an IRAS point source toward Complex H tentatively identified 
as a possible  young stellar object  \citep{ivezic97}, and 
 weak $H_\alpha$ emission toward the brightest part of the Complex 
(Putman et al, in press), but the $H_\alpha$  velocity is $30$ km s$^{-1}$ 
more positive than the velocity of the core and does not correspond
to any feature in \hi.

\section{Summary Comments}

The model  derived from the  kinematics of Complex H 
 is consistent with the overall structure of the object.  
The large-scale ``head-tail'' morphology seen in Fig.~1  
(the trend continues to even more positive velocities than are shown) 
is the form one would expect from a cloud containing dense 
concentrations moving on an inclined, retrograde orbit to 
 higher $\ell$ and lower $b$, shedding gas through interaction 
with the Milky Way.  A number of high-velocity clouds believed to 
be  interacting with the Milky Way also have this structure, 
where, like Complex H, the velocity of the diffuse gas 
``downstream'' in the tail is closer  to Galactic rotation 
than the gas in the head  \citep{pietz,bruns}.  
The physical issues involved in the interaction of a
 high-velocity cloud with an  external medium have stimulated considerable 
theoretical interest  \citep{benjamin99}; magnetic fields may 
be important in preserving the  integrity of the main cloud 
\citep{santillan,konz,quilis}. 

In its small-scale structure Complex H resembles 
parts of the Magellanic Stream.  
Though its origin is tidal, the Stream shows evidence for 
 interaction with an external medium \citep{putman98,konz}. 
It contains  \hi\ in two phases, and clouds with  a size of 15--30 pc, 
$\langle n \rangle \approx 2-6$ cm$^{-3}$, and $\Delta v 
\approx 5$ km s$^{-1}$, virtually identical to the knots 
in Complex H \citep{wakk02}.   
In its large-scale characteristics, Complex H  resembles the 
 dwarf irregular galaxy Leo A.    
At a distance of 0.8 Mpc it has about the same \hi\ mass as  the Complex 
and the same linear size as the core; the  \hi\ lines 
from the two objects have a similar width and  
two-component structure \citep{younglo96,schulte}.  
Most dwarf irregular galaxies, however, 
 have considerably more \hi\ mass than Complex H \citep{mateo}.
The Complex  resembles as well  some ``compact'' high-velocity 
\hi\ clouds which,  if they are at distances of several hundred kpc, 
have a similar size and \hi\ mass and also have two \hi\ phases 
\citep{burton01,deHeij02}.  

Complex H may be interacting with the Galactic halo 
 or some part of the  disk.  Its orbital period 
for $V_c=220$ km s$^{-1}$ is $\sim 10^9$ yr over which time it 
traverses heights of $\pm 25$ kpc from the Galactic plane. 
There is  convincing evidence for a gaseous  halo 
that  extends to at least  $R \sim 50$ kpc toward the Magellanic Stream,
 though little is known of it in detail 
\citep{savage95,kalberla_kerp,  konz, savage03}. The 
normal \hi\  disk  emission at $\ell \approx 130\arcdeg$ goes to 
$V_{LSR} \approx -110$ km s$^{-1}$  which, for a flat rotation curve, 
is $R=25$ kpc, similar to the  derived distance of the Complex.  
The Galaxy may also 
have an extended disk of ionized gas like some other systems 
\citep{blandhaw97}.  The existing GBT \hi\ data 
do not cover enough area to be useful in evaluating the suggestion of 
\citet{morras} that Complex H has actually pierced the \hi\ disk, but  
further  observations  are planned.  

Galaxies exist in a rich environment of companions, interaction and 
distortion \citep{sancisi99a}.
Retrograde motions are observed in \hi\ components of 
many systems \citep{bertola}, 
and in Milky Way  globular clusters  \citep{freeman02}. 
 A retrograde orbit 
at large R tends to be less destructive to a satellite  galaxy 
than a prograde orbit \citep{tt,hernquist}, 
which may account for the persistence of the object.  
But Complex H may not persist much longer.  If a 
 complete accounting of the mass of  diffuse, presumably 
decelerated, gas associated with the Complex could be made, it
might well show that the coherent core is only a fraction of 
the total \hi, implying that Complex H is now in the 
process of being destroyed.

Finally, the guiding assumption 
that Complex H is in  a purely circular orbit  (or is just now at 
perigalacticon) is clearly too naive --- 
there is likely some ellipticity and a 
radial velocity component. Additional data will supply extra constraints 
on $V_{\theta}$.  
But the simple model presented here,
motivated by the new evidence for  interaction with the Galaxy, 
suggests that the velocity of Complex H and its gradient with 
latitude carry significant information about its space motion, and 
are not  entirely contingent.

\acknowledgments

I thank R.A.~Benjamin, W.B.~Burton, S. Majewski, E.M.~Murphy, 
T.M.~Tripp, and the referee, H. van Woerden, for useful discussions and 
comments on the manuscript.

% the tables

% the figures

\clearpage
\begin{figure}
\epsscale{0.85}
\plotone{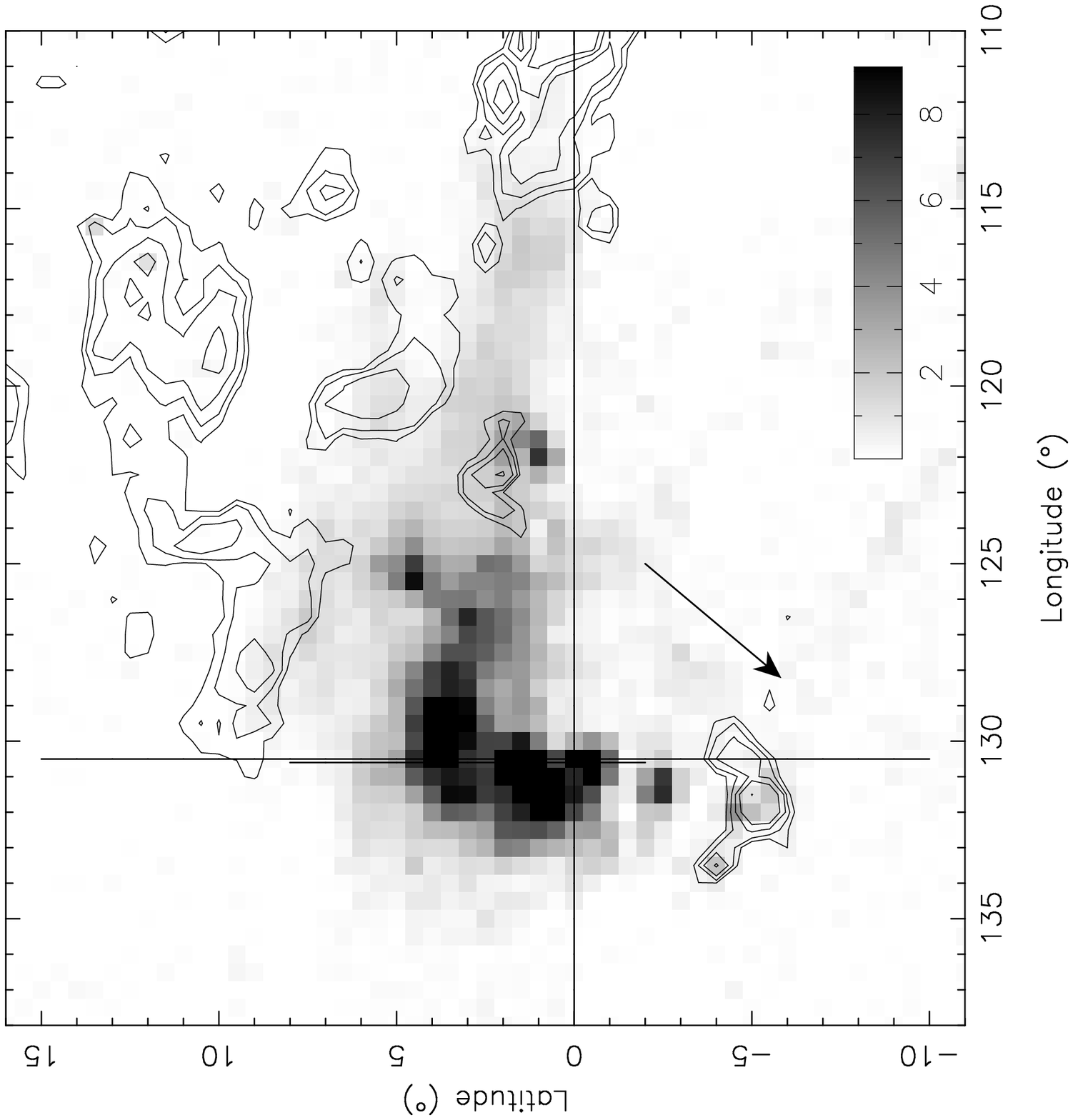}
\caption{Complex H in the Leiden-Dwingeloo 
21cm \hi\ survey data \citep{hartmann} at $35\arcmin$ angular resolution.
The shading shows  \hi\ at 
 $V_{LSR} \leq  -175$ km s$^{-1}$, which  includes the 
brightest \hi\ in the Complex.  The contours trace 
 the \hi\ emission at more positive velocities 
between $-175$ and $-140$ km s$^{-1}$, closer to that allowed by 
normal Galactic rotation.  
The vertical lines near $\ell = 130\arcdeg$ show the locations of 
 the  velocity-latitude cuts of Figs. 2 and 3, which were used 
to derive the model for the cloud's motion 
in the direction shown by the arrow. 
In this model, most of the contoured material would have been decelerated 
by interaction with the Milky Way. The grey scale is in units of 
$10^{19}$ cm$^{-2}$; coutours are equally spaced every $2 \times 10^{19}$ 
cm$^{-2}$.
}
\end{figure}
\clearpage

\begin{figure}
\epsscale{0.65}
\plotone{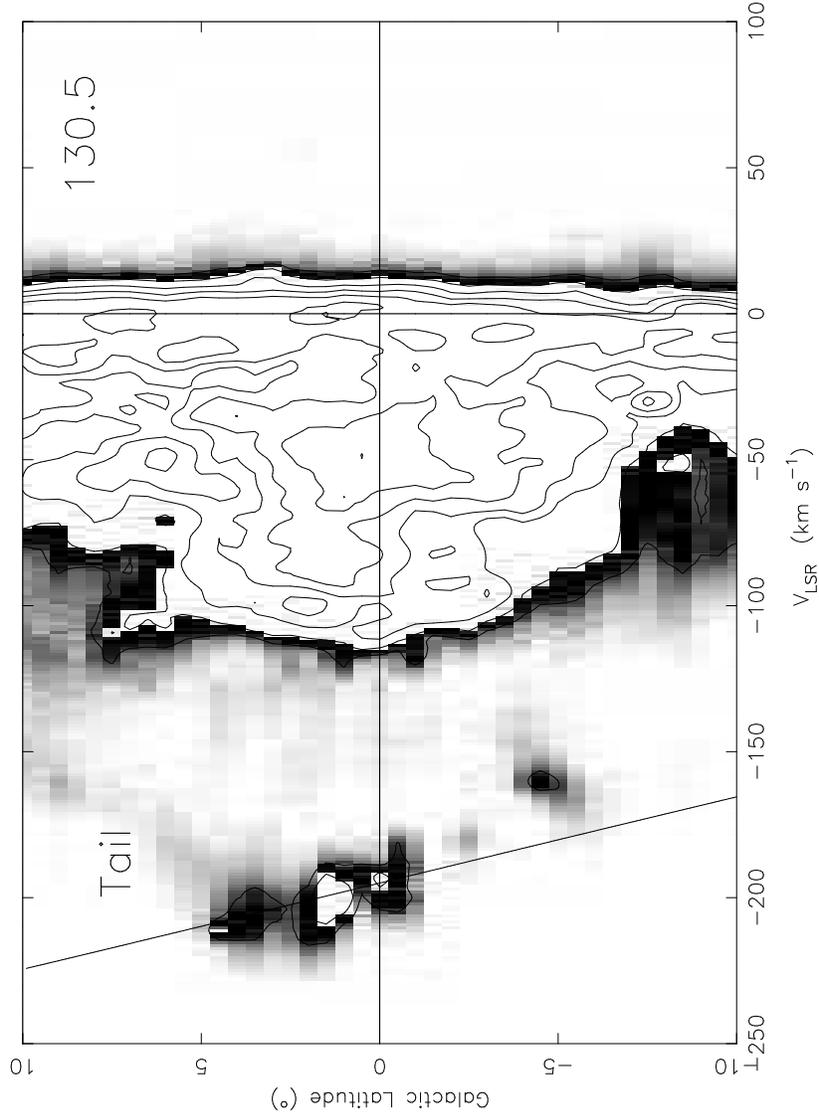}
\caption{A velocity-latitude cut 
(a ``slit spectrum'') through Complex H at  
$\ell = 130\fdg5$ from the LD survey data showing the 
 ``core'' at $b<5\arcdeg$ and the ``tail'' at $b>5\arcdeg$.  
 The slanting line through the core 
follows $V_{LSR}=-195 \ -170\  \sin(b)$ 
km s$^{-1}$.  The slope of the line gives the  vertical component of 
orbital motion, $V_z$, while its value at $b=0\arcdeg$ constrains 
$V_\theta/R$.    Contours are in factors of two 
from the peak; the gray scale shows emission at $T_b \leq 3$ K.  
}
\end{figure}

\clearpage
\begin{figure}
\epsscale{1.3}
\plottwo{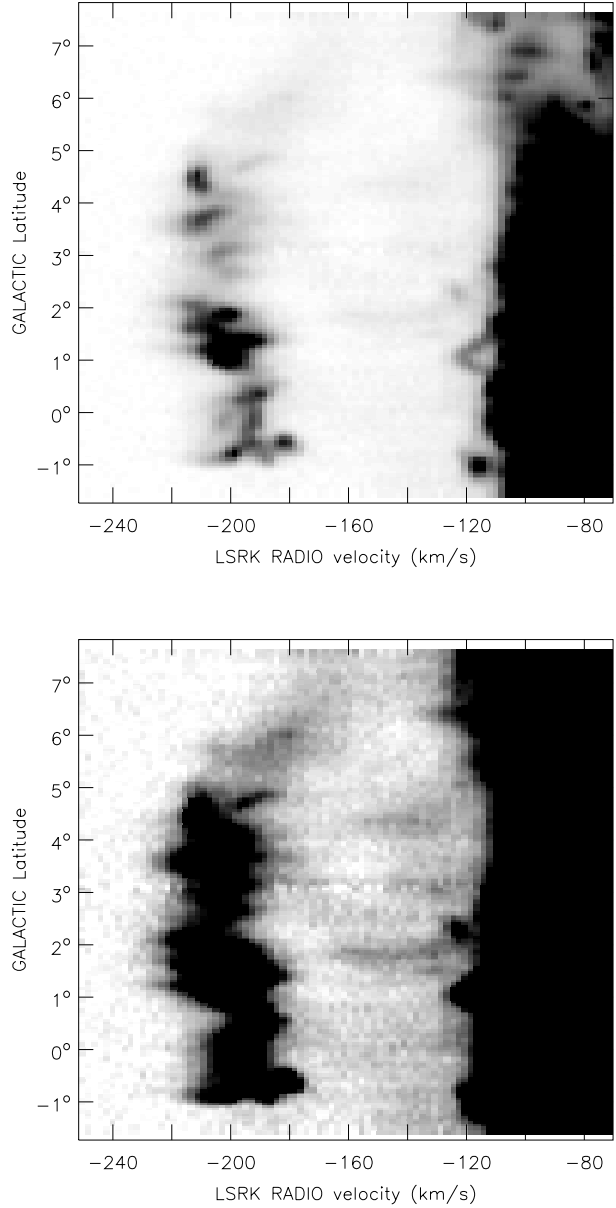}{f3b.eps}
\caption{A velocity-latitude cut through Complex H at 
$\ell = 130\fdg6$ made from GBT data at $9\arcmin$ angular resolution.  
The two panels show identical data; the upper panel is 
clipped at 5 K, the lower at 1 K.  The ``tail'' at $b > 5 \arcdeg$ 
is almost entirely broad-line gas while the bright ``core'' at $b < 
5\arcdeg$ contains many knots of narrower lines. There is  diffuse 
\hi\ emission at velocities between 
Complex H and the Galactic disk suggesting that some material 
from the Complex has been decelerated by interaction with 
the Milky Way.  
}
\end{figure}

\clearpage
\begin{figure}
\epsscale{0.65}
\plotone{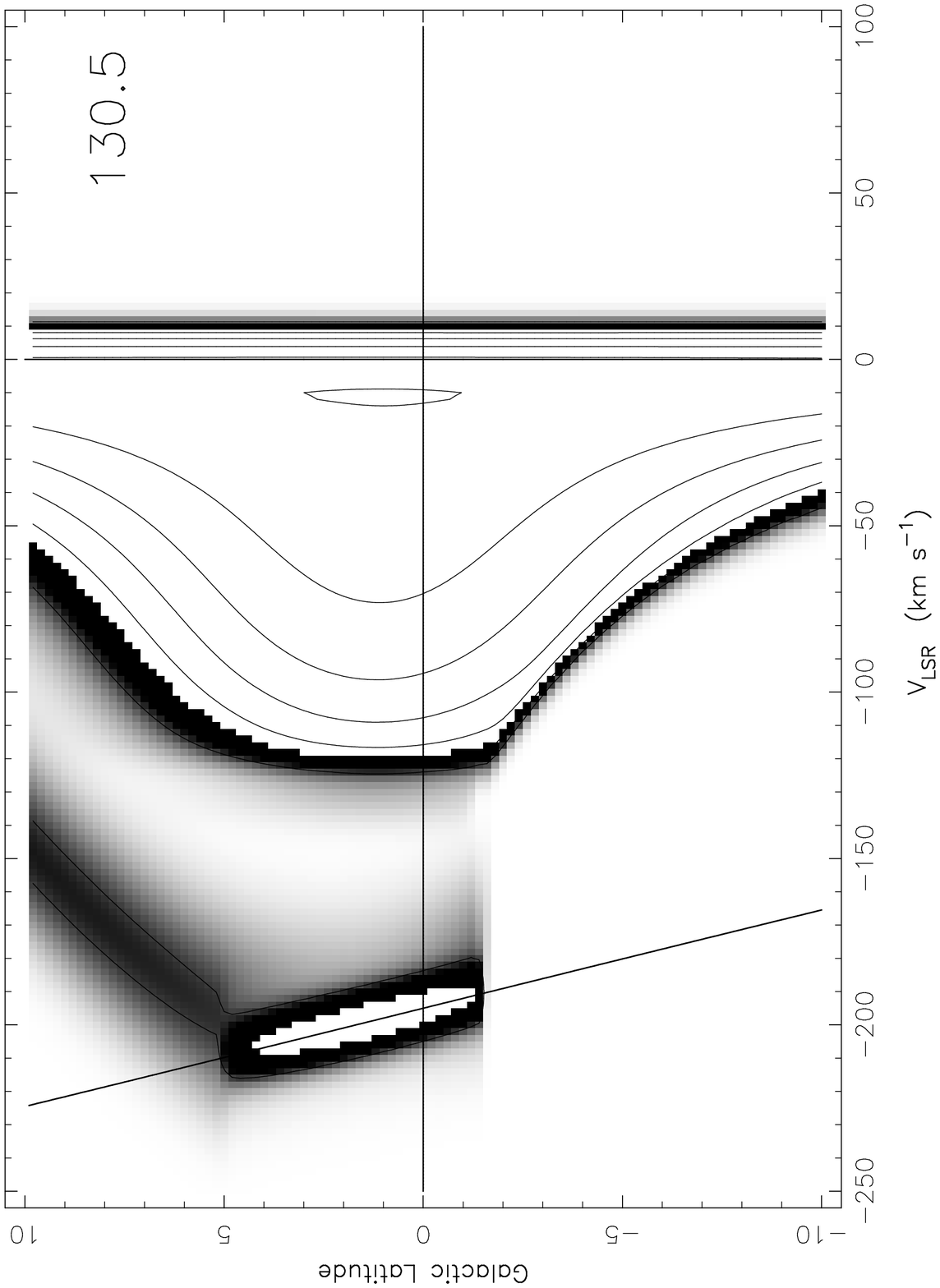}
\caption{Simulation of Complex H as a cloud in an inclined retrograde orbit
 surrounded by a diffuse component whose velocity changes linearly between
 the retrograde core and normal prograde Galactic rotation.
The simulation is intended to match the observations of Figs.~2 and 3. 
The velocity tilt of the core results from the projection of 
its vertical motion.  
The diagonal line is the same one drawn through the data in Figure 2.
}
\end{figure}


\begin{thebibliography}{}

\bibitem[Benjamin (1999)]{benjamin99} Benjamin, R.A. 1999, in 
Stromlo Workshop on High Velocity Clouds, ASP Conf. Ser. Vol. 166, p. 147

\bibitem[Bertola \& Corsini (1998)]{bertola} Bertola, F., \& Corsini, E. 
1998, in Galaxy Interactions at Low and High Redshift, IAU Symp. 186, 
ed. J.E. Barnes \& D.B. Sanders, Kluwer, p. 149
%% Counterrotation in Galaxies

\bibitem[Bland-Hawthorn, Freeman, \& Quinn (1997)]{blandhaw97} 
  Bland-Hawthorn, J., Freeman, K.C., \& Quinn, P.J. 1997, \apj, 490, 143
%%Where do the disks of spiral galaxies end?  H-alpha extensions

\bibitem[Blitz et al. (1999)]{blitz} Blitz, L., Spergel, D.N., Teuben, P., 
Hartmann, D., \& Burton, W.B. 1999, \apj, 514, 818
%% hvcs as members of the local group

\bibitem[Br\"{u}ns et al. (2000)]{bruns} Br\"{u}ns, C., Kerp, J., Kalberla, 
P.M.W., \& Mebold, U. 2000, \aap, 357, 120
%%% looking for head-tail hvcs

\bibitem[Burton, Braun, \& Chengalur (2001)]{burton01} Burton, W.B., 
Braun, R., \& Chengalur, J.N. 2001, \aap, 375, 227
%% Arecibo imaging of compact high-velocity clouds

\bibitem[Cram \& Giovanelli (1976)]{cram} Cram, T.R., \& Giovanelli, R. 
1976, \aap, 48, 39
%% Two-component structure in the profiles of high velocity clouds

\bibitem[Dieter (1971)]{dieter} Dieter, N.H. 1971, \aap, 12, 59

\bibitem[Ferrara \& Field (1994)]{ferrara94} Ferrara, A., \& Field, G.B. 1994, ApJ, 423, 665
% pressure confinment of hvcs -- two phase model

\bibitem[Freeman \& Bland-Hawthorn (2002)]{freeman02} Freeman, K., \& 
Bland-Hawthorn, J. 2002, \araa, 40, 487

\bibitem[Hartmann \& Burton (1997)]{hartmann} Hartmann, D. \& Burton, W.B., 
1997, ``Atlas of Galactic Neutral Hydrogen'', Cambridge University Press 

\bibitem[de Heij, Braun, \& Burton (2002)]{deHeij02} de Heij, V., Braun, R., 
  \& Burton, W.B. 2002, \aap, 391, 67

\bibitem[Hernquist \& Quinn (1988)]{hernquist} Hernquist, L. \& Quinn, P.J. 1988, \apj, 331, 682

\bibitem[Hulsbosch (1971)]{huls71} Hulsbosch, A.N.M. 1971, \aap, 14, 489

\bibitem[Hulsbosch (1975)]{huls75} Hulsbosch, A.N.M. 1975, A\&A, 40, 1

\bibitem[Ivezi\'{c} \& Christodoulou (1997)]{ivezic97} Ivezi\'{c}, \v{Z}., \& 
	Christodoulou, D.M. 1997, \apj, 486, 818
%% IR search for young stars in HI high-velocity clouds

\bibitem[Kalberla \& Kerp (1998)]{kalberla_kerp} Kalberla, P.M.W., \& 
Kerp, J. 1998, \aap, 339, 745 
%% Hydrostatic equilibrium consditions in the galactic halo

\bibitem[Konz, Br\"{u}ns, \& Birk (2002)]{konz} Konz, C., Br\"{u}ns, C., 
\& Birk, G.T. 2002, A\&A, 391, 713

\bibitem[Lockman (2002)]{fjl2002} Lockman, F.J. 2002, ApJ, 580, L47

\bibitem[Mateo (1998)]{mateo} Mateo, M. 1998, \araa, 462, 203
%% review article on dwarf galaxies of the local group

\bibitem[Morras, Bajaja, \& Arnal (1998)]{morras} Morras, R., Bajaja, E., \& Arnal, E.M. 1998, A\&A, 334, 659
%% Complex H: A case of HVC-galaxy collision

\bibitem[Pietz et al  (1996)]{pietz} Pietz, J., Kerp, J., Kalberla, P.M.W., 
Mebold, U., Burton, W.B. \& Hartmann, D. 1996, A\&A, 308, L37

%\bibitem[Putman (2000)]{putman00} Putman, M.E. 2000, Pub. Astr. Soc. 
%Australia, 17, 1
%% structures in the mag stream

\bibitem[Putman et al (1998)]{putman98} Putman, M.E. et al 1998, Nature, 
394, 752
%% Tidal disrution of the Magellanic Clouds by the Milky Way
%% detection of the leading arm

\bibitem[Quilis \& Moore (2001)]{quilis} Quilis, V., \& Moore, B. 2001, 
  \apj, 555, L95
%% Where are the high-velocity clouds?  hydro models 

\bibitem[Sancisi (1999)]{sancisi99a} Sancisi, R. 1999, in 
 Galaxy Interactions at Low and High Redshift, IAU Symposium 
186, ed. J.E. Barnes \& D.B. Sanders, Kluwer, p. 71
%% Review: Galaxy Interactions: the HI signature

\bibitem[Santillan et al (1999)]{santillan} Santillan, A., Franco, J., 
Martos, M., \& Kim, J. 1999, \apj, 515, 657.

\bibitem[Savage (1995)]{savage95} Savage, B.D. 1995, in The Physics 
of the Interstellar Medium and Intergalactic Medium, ed. A. Ferrara, 
C.F. McKee, C. Heiles, \& P.R. Shapiro, ASP Conf. Ser. 80, p. 233
%% The Galactic Corona -- good review

\bibitem[Savage et al (2003)]{savage03} Savage, B.D., Sembach, K.R., 
Wakker, B.P., Richter, P., Meade, M., Jenkins, E.B., Shull, J.M., 
Moos, H.W., \& Sonneborn, G. 2003, \apjs, 146, 125

\bibitem[Schulte-Ladbeck et al (2002)]{schulte} Schulte-Ladbeck, R.E., 
Hopp, U., Drozdovsky, I.O., Greggio, L., \& Crone, M.M. 2002, \aj, 124, 896
%% distance to Leo A

\bibitem[Toomre \& Toomre (1972)]{tt} Toomre, A., \& Toomre, J. 1972, 
ApJ, 178, 623
%% Galactic bridges and tails

\bibitem[Wakker, Vijfschaft, \& Schwarz (1991)]{WakkVS} Wakker, B.P., 
Vijfschaft, B., \& Schwarz, U.J. 1991, A\&A, 249, 233

\bibitem[Wakker \& Schwarz (1991)]{WS91} Wakker, B.P., \& Schwarz, U.J. 1991, A\&A, 250, 484

\bibitem[Wakker et al. (1998)]{wakker98} Wakker, B.P., van Woerden, H., 
de Boer, K., \& Kalberla, P. 1998, ApJ, 493, 762.
%% A lower limit to the distance of the high-velocity cloud
%% Complex H

\bibitem[Wakker, Oosterloo, \& Putman (2002)]{wakk02} Wakker, B.P., Oosterloo, 
T.A., \& Putman, M.E. 2002, \aj, 123, 1953

\bibitem[Young \& Lo (1996)]{younglo96} Young, L. \& Lo, K.Y. 1996, \apj, 
	462, 203
%% The neutral ISM in dwarf galaxy Leo A.

\bibitem[Zaritsky (1999)]{zaritsky} Zaritsky, D. 1999, in The Third 
Stromlo Symposium: The Galactic Halo, ed. B.K. Gibson, T.S. Axelrod, M.E.
Putman, ASP Conf. Ser. Vol. 165, 34
%% The Mass and extent of the Galactic Halo

\end{thebibliography}
\end{document}